\documentclass[english,aps,manuscript,journal=mamobx,manuscript=article,maxauthors=15,biblabel=plain]{revtex4}
\usepackage[T1]{fontenc}
\usepackage[latin9]{inputenc}
\usepackage[a4paper]{geometry}
\usepackage{amsthm}
\usepackage{amsmath}
\usepackage{amssymb}
\usepackage{graphicx}

\makeatletter
\@ifundefined{textcolor}{}
{%
 \definecolor{BLACK}{gray}{0}
 \definecolor{WHITE}{gray}{1}
 \definecolor{RED}{rgb}{1,0,0}
 \definecolor{GREEN}{rgb}{0,1,0}
 \definecolor{BLUE}{rgb}{0,0,1}
 \definecolor{CYAN}{cmyk}{1,0,0,0}
 \definecolor{MAGENTA}{cmyk}{0,1,0,0}
 \definecolor{YELLOW}{cmyk}{0,0,1,0}
}
\numberwithin{equation}{section}
\numberwithin{figure}{section}

\makeatother

\usepackage{babel}
\begin{document}

\title{The effect of topology on the conformations of ring polymers}

\author{M. Lang, J. Fischer, and J.-U. Sommer}
\begin{abstract}
The bond fluctuation method is used to simulate both non-concatenated
entangled and interpenetrating melts of ring polymers. We find that
the swelling of interpenetrating rings upon dilution follows the same
laws as for linear chains. Knotting and linking probabilities of ring
polymers in semi-dilute solution are analyzed using the HOMFLY polynomial.
We find an exponential decay of the knotting probability of rings.
The correlation length of the semi-dilute solution can be used to
superimpose knotting data at different concentrations. A power law
dependence $f_{n}\sim\phi R^{2}\sim\phi^{0.77}N$ for the average
number $f_{n}$ of linked rings per ring at concentrations larger
than the overlap volume fraction of rings $\phi^{*}$ is determined
from the simulation data. The fraction of non-concatenated rings displays
an exponential decay $P_{OO}\sim\exp(-f_{n})$, which indicates $f_{n}$
to provide the entropic effort for not forming concatenated conformations.
Based upon this results we find four different regimes for the conformations
of rings in melts that are separated by a critical lengths $N_{OO}$,
$N_{C}$ and $N^{*}$. $N_{OO}$ describes the onset of the effect
of non-concatenation below which topological effects are not important,
$N_{C}$ is the cross-over between weak and strong compression of
rings, and $N^{*}$ is defined by the cross-over from a non-concatenation
contribution $f_{n}\sim\phi R^{2}$ to an overlap dominated concatenation
contribution $f_{n}\sim\phi N^{1/2}$ at $N>N^{*}$. For $N_{OO}<N<N_{C}$,
the scaling of ring sizes $R\sim N^{2/5}$ results from balancing
non-concatenation with weak compression of rings. For $N_{C}<N<N^{*}$,
non-concatenation and strong compression imply $R\sim N^{3/8}$. Our
simulation data for non-interpenetrating rings up to $N=1024$ are
in good agreement with the prediction for weakly compressed rings.
\end{abstract}
\maketitle

\section{Introduction}

Even though ring polymers have an extremely simple structure similar
to linear chains but without ends, there are still many unsolved questions
concerning their properties in concentrated solutions or melts. The
main reason for the difficulty to describe ring polymers lies in the
fixed topological state of each individual ring and the topological
interactions with overlapping molecules. The topology and classification
of individual rings in terms of ``knots'' is a long standing problem
in science ranging from early discussions on the nature of atoms \cite{Thomson}
and mathematics \cite{Adams} to applications in modern science \cite{Orlandini2007}.
This is an important point, since for polymers it is known, for instance,
that the knot type affects swelling, collapse and average size of
a ring polymer \cite{Moore_Grosberg_2005,Grosberg_Feigel_Rabin}.

Overlapping rings can also form permanently entwinned states or ``links''.
It was suggested that giant aggregates of such entwinned rings could
form based on topological connectivity only, so called Olympic gels
\cite{Degennes,raphael1997,pickett2006}, which might allow to directly
measure the entanglement contribution to elasticity. Simulations of
melts of rings showed that topological interactions mutually compress
the conformations of overlapping rings \cite{Muller_1996,BrownSzamel98,Vettorel2009,Rosa,Tsolou}.
Cates and Deutsch (CD) conjectured that this compression might lead
to a scaling of ring size $R_{g}\sim N^{2/5}$ somewhere in between
fully collapsed and ideal conformations \cite{cates1986}. Such a
behavior is possible, since in melts, excluded volume interactions
are typically screened and thus, topological constraints, which are
often incorporated into the effects of excluded volume, become important.
Therefore, the effect of topology, unlike the excluded volume effects
in other situations, does not lead to a simplification or idealization
of ring conformations, when bringing rings into overlap. The basic
argument in the CD-model is that all overlapping rings which could
interpenetrate in the absence of topological constraints are rejected
from the volume of gyration of an individual ring in the case of non-interpenetration.
The corresponding free energy effort is balanced by the free energy
penalty arising from confining an ideal linear chain without taking
into account topology. Recent simulation studies suggest that the
conjecture of CD is only characteristic of an intermediate regime
before the scaling of ring size crosses over to a compact state $R_{g}\sim N^{\nu}$
with $\nu$ close to 1/3 as summarized in Fig. 1 of Ref. \cite{Halverson}.
A similar cross-over has been predicted recently \cite{Sakaue2011}
by proposing a model that needs to make assumptions on the nature
of the knotting and the non-concatenation constraint that enter in
the free energy of a polymer ring in melt. 

It is the goal of the present work to clarify the assumptions made
in the above works concerning non-concatenation and knotting for a
better understanding of the effect of topology onto the conformations
of rings in melt. In the present work, we study both solutions of
entangled non-concatenated rings and interpenetrating rings spanning
the regime between dilute to melt concentrations using the bond fluctuation
method. We analyze conformations in both types of ring-solutions and
determine the linking and knotting probabilities as a function of
concentration and degree of polymerization. These results are then
combined with previous work on knotting of rings \cite{Grosberg_Feigel_Rabin,Moore_Grosberg_2005}
in order to estimate the conformations of on rings in melts of rings
and linear chains.

\section{\label{sec:Methods}Simulation Methods and Simulated Systems}

We use the bond-fluctuation model (BFM) \cite{CarmesinKremer88} to
simulate solutions of ring polymers. This method was chosen, since
it is known to reproduce conformational properties and dynamics of
semi-dilute solutions \cite{paul1991crossover} and polymer networks
\cite{Sommer:Bfm1,Lang2010,lang2003length}. In this method, each
monomer is represented by a cube occupying eight lattice sites on
a cubic lattice. In the standard definition of this algorithm, the
bonds between monomers are restricted to a set of 108 bond vectors
which ensure cut-avoidance of polymer strands by checking for excluded
volume. Monomer motion is modeled by random jumps to one of the six
nearest lattice positions. A move is accepted, if the bonds connecting
to the new position are still within the set of bond vectors and if
no monomers overlap. All samples of the present study were created
in simulation boxes with periodic bondary conditions. Athermal solvent
is treated implicitly by empty lattice sites.

In the present work, we use this method to create solutions of non-concatenated
rings. Additionally, we perform simulations where we allow for moves
diagonal to the lattice axis. This switches off all entanglements,
while excluded volume interactions are mainly unaffected. In consequence,
the rings can interpenetrate each other to form concatenated conformations.
In order to clarify the differences between both series of samples,
we call them \emph{interpentrating} and \emph{non-interpenetrating}
throughout this work. In both cases, relaxation of the rings was monitored
by the autocorrelation function of the vectors connecting opposite
monomers of a ring. All samples were relaxed several relaxation times
of the ring polymers. Afterward, chain conformations were analyzed
from snapshots of the ring solutions from a very long simulation run.
The error of the data points was computed by the total number of statistically
independent conformations available.

For the analysis of topology, monomer motion in the interpenetrating
samples was slowly switched back to motion along the lattice axis
in order to remove states in which pairs of bond vectors are intersecting
each other. Then, rings were first simplified at conserved topology
and regular projections of any pair of overlapping rings were determined
as described in \cite{lang2003analysis}. Finally, the Gauss code
of the projection was computed and the types of knots and pairwise
links formed were analyzed using the Skein-Template algorithm of Gouesbet
\emph{et al.} \cite{Gouesbet99} based upon the HOMFLY-polynomial
\cite{HOMFLY}. Note that in the present paper we restricted the analysis
to permanent entanglements between \emph{pairs} of rings. Brunnian
links (permanent entanglements that cannot be detected by a pairwise
analysis of rings) as for instance, the Borromean rings \cite{Adams}
cannot be detected by this approach. The data of previous work \cite{Michalke2001}
shows that this simplification should affect the results of the present
study by less than 5\%, which is in the range of the accuracy of the
individual data points. 

We ran an array of different interpenetrating and non-interpenetrating
samples with degrees of polymerization $N=$16, 32, 64, 128, 192,
256, 384, 512, 768, and 1024 and polymer volume fractions of $\phi=0.5$,
0.375, 0.25, 0.1875, 0.125, 0.0625, and 0.03125. The numbers of rings
per sample varied from 512 to 4096 resulting in a simulation box size
between $128^{3}$ to $512^{3}$ lattice sites.

\section{Conformations of Ring Polymers}

The conformations of ideal (no excluded volume and no entanglements)
rings of $N$ segments are fully described by random walks that return
to the origin. The situation is more complex for real ring polymers,
since the fixed topological state of the molecules \cite{Grosberg_Feigel_Rabin}
and excluded volume affect the statistical properties of the ring
conformations. However, the main properties of excluded volume for
linear polymers are expected to be valid also for unknotted ring polymers
\cite{Casassa}, since the length at which knotting becomes important
is typically large \cite{Moore_Grosberg_2005} as we also show in
section \ref{sec:Knotting-and-linking}.

Let us define the overlap volume fraction as the volume fraction of
a polymer in its own radius of gyration in good solvent $R_{g_{0}}$
according to 
\begin{equation}
\phi^{*}=N/R_{g_{0}}^{3}.\label{eq:phistar}
\end{equation}
At polymer volume fractions $\phi\gtrsim\phi^{*}$, both interpenetrating
and non-interpenetrating rings mutually compress each other. For interpenetrating
rings, compression is driven by screening of excluded volume effects
and a corresponding change of the blob size $\xi$. For non-interpenetrating
rings topological interactions (exclusion of interpenetrating states)
have to be taken in into account. Let 
\begin{equation}
R_{g_{0}}\approx bN^{\nu_{0}}\ \ \ \ \mbox{for}\ \phi\ll\phi^{*}\label{eq:R0}
\end{equation}
with $\nu_{0}\approx0.587597(7)$ \cite{Clisby} denote the size of
the swollen rings in good solvent and 
\begin{equation}
R_{g}\approx bN^{\nu}\ \ \ \ \mbox{for}\ \phi\gg\phi^{*}\label{eq:Rg}
\end{equation}
is the size of rings in semi-dilute solution. Since the overlap-concentration
is related to the onset of both excluded volume and pairwise topological
effects, a scaling is expected according to 
\begin{equation}
\frac{R_{g}(\phi)}{R_{g_{0}}}\sim f\left(\frac{\phi}{\phi^{*}}\right)\,\,.\label{eq:Rgc}
\end{equation}
The asymptotic limits of the scaling function, $f(y),$ can be obtained
using Eqs.(\ref{eq:phistar})-(\ref{eq:Rg}) as follows:
\begin{equation}
f(y)=\left\{ \begin{array}{c}
1\\
y^{\frac{\nu_{0}-\nu}{1-3\nu_{0}}}
\end{array}\right.\ \ \ \ \mbox{for}\ \ \begin{array}{c}
\phi\ll\phi^{*}\\
\phi\gg\phi^{*}
\end{array}\label{eq:function}
\end{equation}
Different values for the exponent $\nu$ are suggested in literature
to describe the size $R\approx bN^{\nu}$ of interpenetrating ($\nu=1/2$)
or non-interpenetrating $(\nu=2/5$, $1/3$) rings in melt \cite{cates1986,Sakaue2011,Vettorel2009}.
Thus, we expect the following power laws, $R\sim\phi^{x}$, for $\phi\gg\phi^{*}$
\begin{equation}
R\sim\begin{cases}
\phi^{-0.115}\ \ \ \mbox{for} & \nu=1/2\\
\phi^{-0.25}\ \ \ \mbox{for} & \nu=2/5\\
\phi^{-0.33}\ \ \ \mbox{for} & \nu=1/3
\end{cases}\label{eq:powers}
\end{equation}
Here, the first case corresponds to interpenetrating rings and is
equivalent with the behavior of linear chains in semi-dilute solutions
\cite{Degennes}. The other two cases correspond to the two suggested
regimes of the non-interpenetrating rings.

\begin{figure}
\includegraphics[angle=270,width=1\columnwidth]{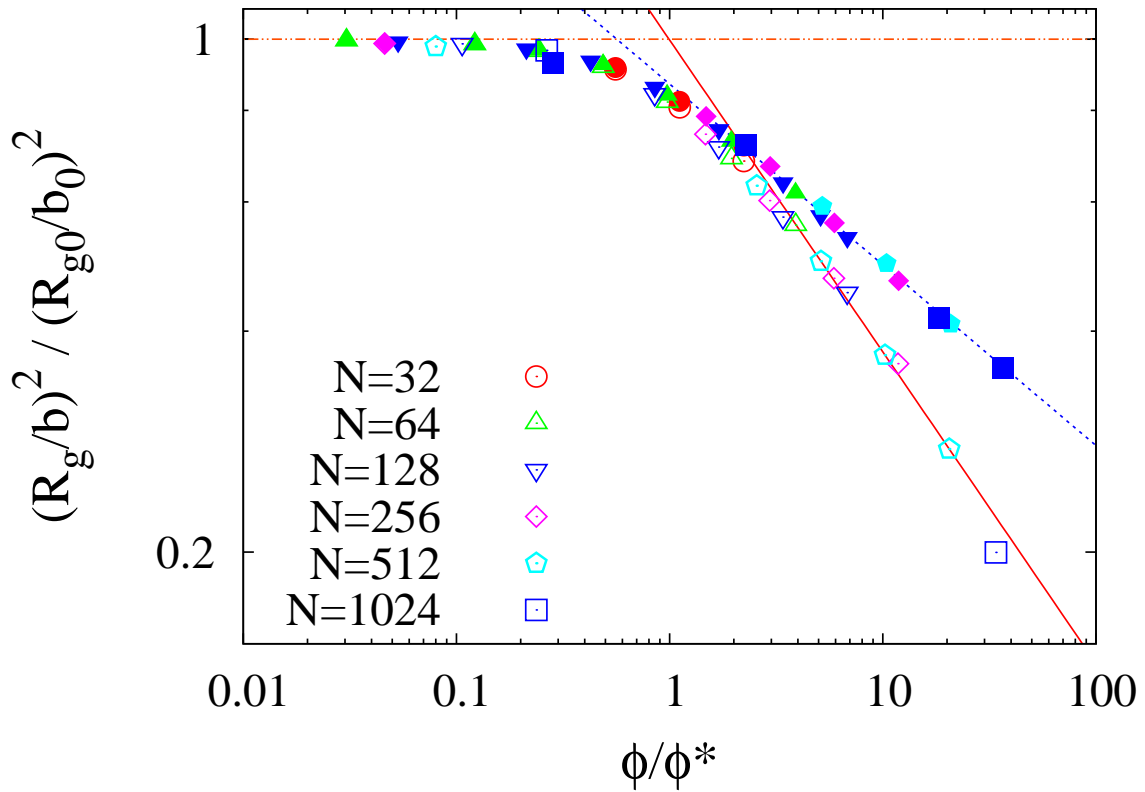}

\caption{\label{fig:Conformations-of-ring}(color online) The size of interpenetrating
(full symbols) and non-interpenetrating (open symbols) ring molecules
in solution. The error of the data points is clearly below symbol
size. The red and blue lines represent fits to the interpenetrating
and non-interpenetrating rings at $\phi\gg\phi^{*}$.}
\end{figure}

In Figure \ref{fig:Conformations-of-ring}, we present the simulation
data both for interpenetrating rings and non-interpenetrating rings.
The interpenetrating rings show an apparent exponent (dashed blue
line) of $x\approx-0.13$ for $\phi>\phi^{*}$, which is close to
the above prediction. The data of non-interpenetrating rings (red
line) show a clearly different exponent of $x\approx-0.21$, which
agrees best with the prediction of CD \cite{cates1986} for the available
molecular data. Based upon the data of Figure \ref{fig:Conformations-of-ring}
one might argue that this could be a continuous transition between
fully swollen and fully compressed conformations. But this rather
narrow regime as function of $\phi/\phi^{*}$ stretches out to about
one decade as function of $N$. Furthermore, a stronger $\phi$ dependence
cannot be excluded for larger $\phi/\phi^{*}$, as indicated by the
data point for the largest rings at the highest concentration. 

\section{Knotting and linking probabilities\label{sec:Knotting-and-linking}}

In this section, we investigate only interpenetrating solutions of
rings. Self-entanglements of individual rings are discussed in terms
of a knotting probability, $1-P_{O}$, while concatenation with overlapping
rings is analyzed by a linking probability, $1-P_{OO}$. The complementary
events are the probabilities of finding an unknotted conformation,
$P_{O}$, or a ring that does not entrap any other ring, $P_{OO}$.

An exponential decrease of the appearance of unknotted conformations
$P_{O}\sim\exp(-N/N_{0})$ with a characteristic and model dependent
``knotting length'' $N_{0}$ was found in previous works \cite{KoniarisMuthukumar1990,Michalke2001,Katritch2000,Grosberg_Rabin_2007,Marcone2006,Moore_Grosberg_2006}
and is a rigorous result for dilute lattice polygons \cite{Orlandini2007}.
This functional form was explained by the abundance of local knots
formed within the range of a small number of bonds along the polymer
as compared to knots that require the global ring conformation to
be knotted. If short knots occur predominantly on a local scale, the
probability of finding no knot then requires that consecutively no
small section of the molecule is knotted and thus the probability
for finding an unknot is exponentially decreasing as function of $N$.

\begin{figure}
\includegraphics[angle=270,width=1\columnwidth]{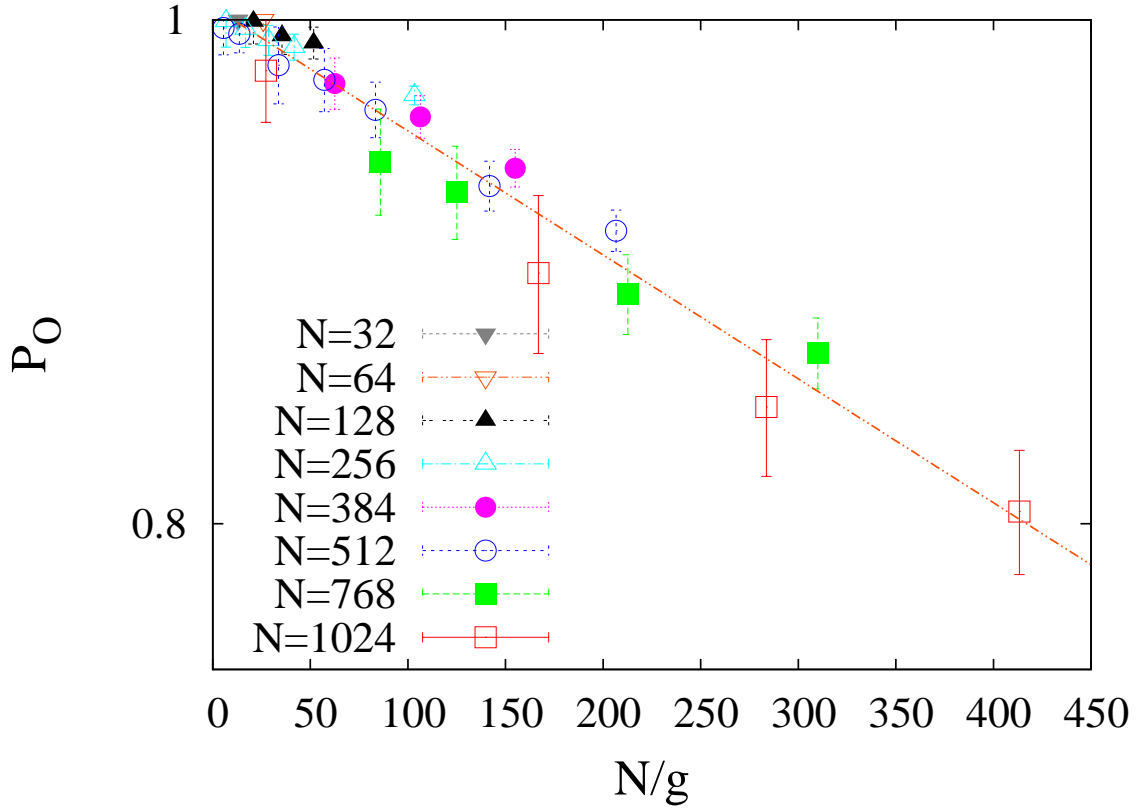}

\caption{\label{fig:Knotting-probability-1}Probability of finding no knot
as function of the number of blobs.}
\end{figure}

Above we showed that for freely interpenetrating rings the same scaling
laws apply as for linear chains in semi-dilute solutions. In the following,
we define the number of monomers per blob, $g$, therefore, as 
\begin{equation}
g=\phi^{-1/(3\nu_{0}-1)}\label{eq:g}
\end{equation}
for simplicity. In the framework of the blob model, it is assumed
that up to the size of the blob, the chains are merely mutually repelling
each other. The blob itself plays thus the role of the basic monomer
unit and the chain or ring on larger scale, composed of $N/g$ blobs,
adopts conformations as in dense melts. If we combine this result
with the above findings on knotting probability, we can postulate
that the simulation data is expected to collapse onto a single curve,
when plotting the probability of finding the ``unknot'' (an unknotted
ring \cite{Adams}), $P_{O}$, as function of the number of blobs
$N/g$. This corresponds to assuming the chain as a tube of diameter
$g^{\nu}$. Figure \ref{fig:Knotting-probability-1} demonstrates,
that this rough estimate nearly leads to a full collapse of the data.
The line is a least squares fit of all data using 
\begin{equation}
P_{O}=\exp\left(-(N/g-a)/N_{0}\right)\label{eq:P_O}
\end{equation}
similar to previous work \cite{Orlandini2007}. The parameter $a=10\pm5$
reflects a minimum required number of blobs to form a knot. We determine
a knotting length $N_{0}=1820\pm80$ for the simulations of the present
study. Note that $N_{0}$ depends strongly on excluded volume: for
random walks one finds $N_{0}\approx340$ \cite{Deguchi,Deguchi2},
while for bead-spring rings with excluded volume one obtains, for
instance $N_{0}\approx8\times10^{5}$ \cite{KoniarisMuthukumar1990}.
Our result is clearly larger than that for the random walk. This might
be due to the incomplete screening of excluded volume on the length
scale of individual monomers \cite{wittmer2007intramolecular}. Since
the knotting length is larger than the largest $N$ of the present
study, we do not expect significant effects of knotting on ring conformations
for the non-interpenetrating samples as long as the rings remain swollen
or close to ideal conformations. Note that there must be an additional
small systematic decrease of $P_{O}$ with increasing $N$ after the
above rescaling of $N$ by $N/g$ as visible in Figure \ref{fig:Knotting-probability-1}.
This is because blobs are soft objects allowing still for some residual
knotting inside the blob. Furthermore, the case $N/g=1$ can be taken
as approximation of isolated self-avoiding walks that exhibit the
above mentioned large knotting length. Both corrections lead to an
additional systematic decrease of the fraction of unknotted rings
at constant $N/g$ for increasing $N$.

\begin{figure}
\begin{center}\includegraphics[width=0.5\columnwidth]{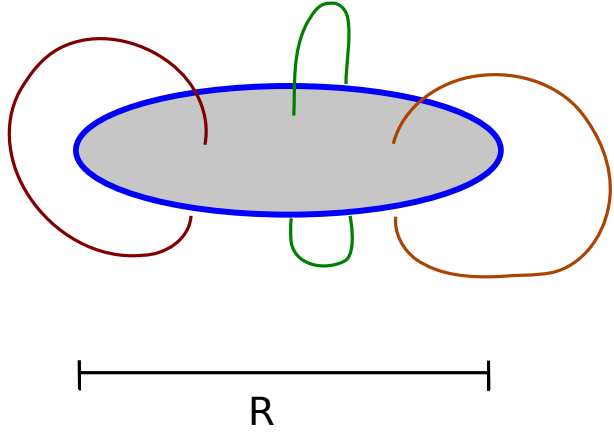}\end{center}

\caption{\label{fig:Model-for-estimating-1-1}Model for estimating the average
number of linked rings.}
\end{figure}

Let us introduce the average number of concatenated rings (linked
rings) per ring, $f_{n}$. In order to estimate $f_{n}$, we first
neglect the effect of multiple concatenations between the same pair
of rings (for instance, link $4_{1}^{2}$ in the standard notation
\cite{Adams}) for sufficiently small $N$ as supported by previous
work \cite{Michalke2001,lang2007trapped}. We now propose that the
number of concatenations per ring is proportional to the minimal surface
$\sim R^{2}$ spanned by the ring %
\footnote{The properties of the minimal surface will be discussed in a forthcomming
work \cite{Langring}.%
} as sketched in Figure \ref{fig:Model-for-estimating-1-1}. The number
of rings in contact with the surface is proportional to the density
$\phi$ of the surrounding rings. Since for semi-dilute solutions
all square sizes (blob and ring) scale as $R^{2}\sim\phi^{-0.23}N$
\cite{Degennes,Rubinstein}, we can approximate 
\begin{equation}
f_{n}\sim\phi R^{2}\sim\phi^{0.77}N\,\,.\label{eq:fn}
\end{equation}

With equation \ref{eq:fn} we assume that a constant fraction of those
ring conformations, that touch the shaded area in Figure \ref{fig:Model-for-estimating-1-1},
leads to a concatenated state. This assumption is reasonable for monodisperse
melts, since because of $N\gg R/b$ essentially all overlapping rings
have the possibility to encircle the other ring independent of their
relative position. This has been demonstrated by the rather constant
linking probabilities of pairs of rings at distances $<R$ in previous
works \cite{lang2004promotion,Michalke2001}.

\begin{figure}
\includegraphics[angle=270,width=1\columnwidth]{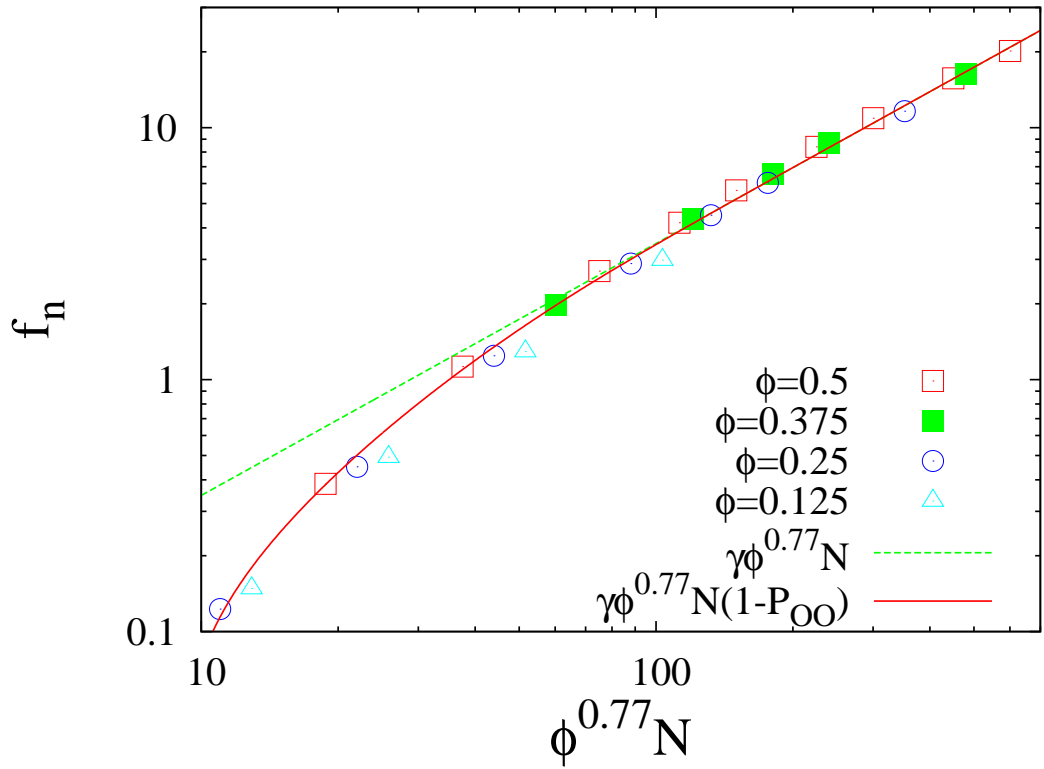}

\caption{\label{fig:Average-numbers-of-1-1}Average number of pairwise concatenations
per ring. Error bars are below symbol size. Lines are theoretical
estimates for $\phi=1.$}
\end{figure}

The average number of concatenations per ring, $f_{n}$, is determined
from the interpenetrating rings and shown in Figure \ref{fig:Average-numbers-of-1-1}.
We find a good agreement with the prediction $\phi^{0.77}N$ of equation
(\ref{eq:fn}) for $N>100$ and a correction for small $N$ that is
discussed below. 

It is important to stress that the average number of entrapped rings
per ring, $f_{n}$, grows quicker with respect to $N$ and $\phi$
than the overlap number of rings, defined as 
\begin{equation}
P=\phi R^{3}/N\approx\phi^{1-3(\nu_{0}-1/2)(3\nu_{0}-1)}N^{1/2}\approx\phi^{0.65}N^{1/2}\,\,.\label{eq:P}
\end{equation}

Nevertheless, we expect a cross-over to an overlap-dominated regime
at a critical degree of polymerization $N^{*}$ at which 
\begin{equation}
f_{n}\approx P.\label{eq:fn2}
\end{equation}
Then, any ring will be concatenated with essentially all overlapping
rings. Figure \ref{fig:Average-numbers-of-1-1} indicates that this
regime is not reached yet. Therefore, the degrees of polymerization
in the present study are all below $N^{*}$ and we can conclude $N^{*}>1024$.

\begin{figure}
\includegraphics[angle=270,width=1\columnwidth]{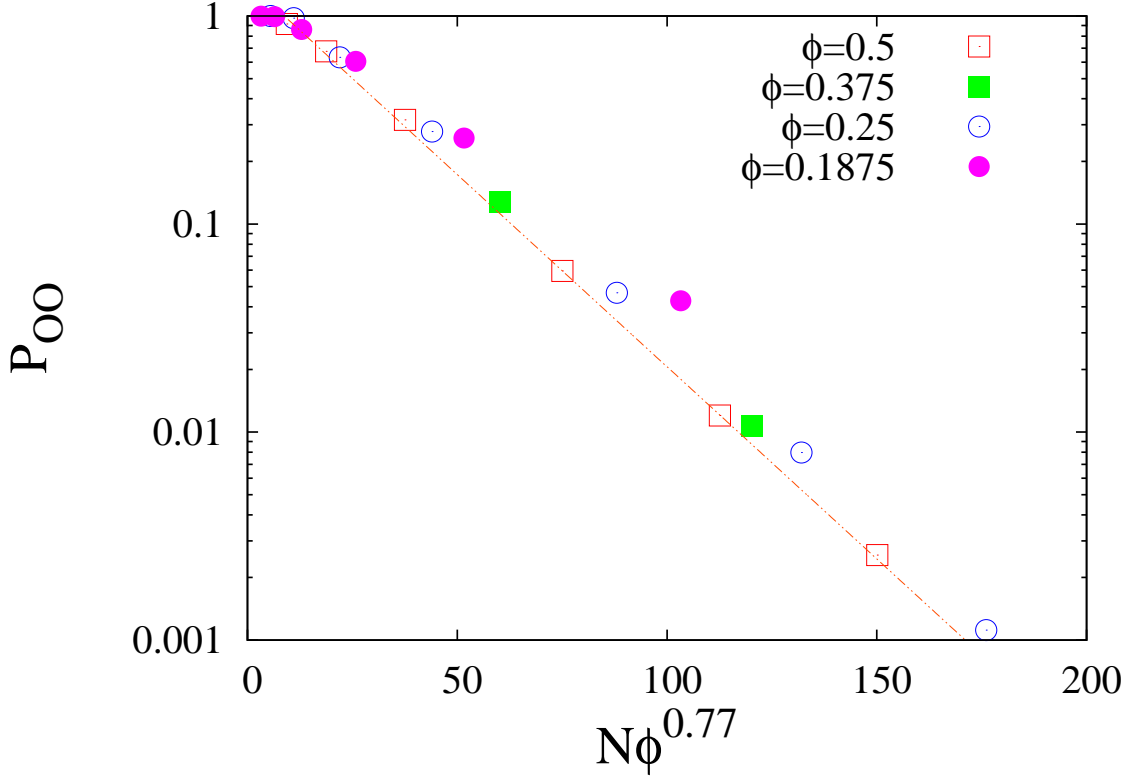}

\caption{\label{fig:Probability-of-finding-1}Probability $P_{OO}$ that a
ring is not linked to any other rings as function of the average number
of linked rings $\phi^{0.77}N$. Error bars are comparable to symbol
size. }
\end{figure}

In a previous work \cite{lang2005network}, an exponential decrease
of $P_{OO}$ as function of $N$ was observed for rings formed in
end-linked model networks. In Figure \ref{fig:Probability-of-finding-1},
we plot $P_{OO}$ as function of $\phi^{0.77}N$ and observe again
a reasonable collapse of the data onto an exponential decay as function
of $N$. The line in Figure \ref{fig:Probability-of-finding-1} is
a least squares fit to the data with $\phi=0.5$ using an expression,
\begin{equation}
P_{OO}=\exp\left(-(N\phi^{0.77}-a)/N_{OO}\right),\label{eq:P_OO}
\end{equation}
similar to the one used for the non-knotting probability above. We
find $a=8.8\pm0.3$ and a linking length $N_{OO}=23.5\pm0.1$ at $\phi=1$
in good quantitative agreement with previous work \cite{lang2005network}. 

Equation (\ref{eq:fn}) for $f_{n}$ can be improved by including
the linking probability $1-P_{OO}$ as shown in Figure \ref{fig:Average-numbers-of-1-1}
by the continuous line. Let us further introduce $\gamma$ as the
constant of proportionality according to $f_{n}=\gamma\text{\ensuremath{\phi}}^{0.77}N(1-P_{OO})$.
From the simulation data in Figure \ref{fig:Average-numbers-of-1-1}
we obtain $\gamma=0.034\pm0.001$. Thus, rings of roughly $50$ monomers
are needed at $\phi=0.5$ to have an average of $f_{n}\approx1$.
For smaller rings, we do not expect a significant effect of topology
onto ring conformations.

It is also instructive to approximate $P_{OO}$ by replacing $N\phi^{0.77}$
by $f_{n}/\gamma$. For large $f_{n}$ we obtain 
\begin{equation}
P_{OO}\approx\exp\left(-f_{n}\right)\label{eq:dhd}
\end{equation}
up to a numerical coefficient of order unity. Thus, the number of
unlinked ring conformations decreases roughly exponential with the
average number of concatenations for sufficiently large $f_{n}$.
This shows that we require a contribution of order $f_{n}$ to the
free energy  to describe the effect of non-concatenation.

Above we mentioned that we neglect Brunnian links. Besides the results
of a previous work \cite{Michalke2001}, the agreement between theoretical
predictions (which ignores whatever type of linking may penetrate
a ring) and our present data (based on pairwise linking) is a strong
indication that restricting the analysis to pairwise linking is a
reasonable approximation for the samples of the present study.

\section{A Flory theory of ring conformations in melt}

With the above results we have the missing data to discuss previous
models \cite{cates1986,Sakaue2011} to explain the scaling of the
size of non-interpenetrating rings. In the former model \cite{cates1986},
CD balance the entropy of confining an ideal ring with the loss of
entropy caused by the exclusion of other polymers from the gyration
volume. This can be expressed in a free energy of form%
\footnote{For simplicity, we drop all (numerical) constants in this section.%
} 
\begin{equation}
\frac{F}{kT}\approx\frac{N}{R^{2}}+\frac{R^{3}}{N}.\label{eq:FECD}
\end{equation}
The optimum size of a ring polymer is then given by 
\begin{equation}
R\sim N^{2/5}.\label{eq:RCD}
\end{equation}
The second term is equivalent to assuming that all potentially overlapping
rings share topological interactions with each other. Our data in
Figure \ref{fig:Average-numbers-of-1-1} show that this is only the
case for molecular weights beyond the transition to the overlap dominated
regime $N>N^{*}$ with $N^{*}>10^{3}$. However, the available simulation
studies \cite{Vettorel2009,Halverson2,Suzuki2009}, including the
results of this work, see Fig.\ref{fig:Conformations-of-ring}, come
to a different conclusion: the predicted exponent, $\nu=2/5$ from
the CD-model agrees better with the simulation data up to $N<10^{3}$.
For larger $N$, on the other hand, there are indications for a change
in the scaling behavior towards $R\sim N^{\nu}$ with $\nu$ close
to $1/3$ \cite{Halverson}. This is clearly surprising and we attempt
to answer this point by a step-wise construction of an alternative
Flory-ansatz for ring conformations in melt.

Moore and Grosberg \cite{Moore_Grosberg_2005} found that the effect
of topology for large unknotted uncompressed rings can be expressed
in terms of an additional excluded volume. This, in contrast to the
native excluded volume, will not be screened in a melt of linear chains
with melt degree of polymerization $>N^{1/2}$. Instead, the topological
contribution stands out and is balanced by the elasticity of the ring.
Let us consider an isolated ring in a melt of linear chains. This
leads to a free energy of form
\begin{equation}
\frac{F}{kT}\approx\frac{R^{2}}{N}+\frac{N^{2}}{R^{3}},\label{eq:Ringlin-1}
\end{equation}
and to a swelling of the ring which is analog to linear chains or
rings in good solvent 
\begin{equation}
R\sim N^{3/5}.\label{eq:Ringlin2-1}
\end{equation}

Let us now return to a monodisperse melt of rings. By comparing the
non-concatenated and the interpenetrating rings we conjecture that
the expulsion of $f_{n}$ rings out of the volume of gyration of a
given ring due to non-concatenation leads to a free energy contribution
$\sim f_{n}\sim R^{2}$ for $N_{OO}<N<N^{*}$. This contribution dominates
by far the elastic free energy of swelling in Equation (\ref{eq:Ringlin-1})
for large $N$. Note that the swelling term is appropriate to balance
ring compression as long as ring size is not strongly compressed,
since then, the $R$-tube confining the ring (see ref. \cite{Grosberg_Feigel_Rabin}
for a detailed discussion) is not yet compressed in length and width.
Thus, for $N_{OO}<N\lesssim N_{C}$ below the crossover to strong
compression at $N_{C}$, the free energy can be approximated by 
\begin{equation}
\frac{F}{kT}\approx\frac{N}{R^{2}}+R^{2}+\frac{N^{2}}{R^{3}}.\label{eq:FET-1-1}
\end{equation}
Here, we formally introduced the entropy loss $N/R^{2}$ upon squeezing
an ideal ring. This term is of minor importance and the free energy
is controlled by balancing two merely topological terms. This results
in an optimal ring size of 
\begin{equation}
R\sim N^{2/5}.\label{eq:size-1}
\end{equation}
Note that our result for the $\nu$-exponent is formally equivalent
to the result by Cates and Deutsch, but is obtained from a different
estimate for non-concatenation for $N<N^{*}$ and a topological contribution
which prevents strong compression of rings at $N<N_{C}$. According
to the available data, for instance \cite{muller1996}, it requires
in melts at $\phi=0.5$ an $N>512$ to compensate the swelling of
the rings due to incompletely screened excluded volume \cite{wittmer2007intramolecular}.
The above scaling of ring size is supported by simulation data of
rings with moderate degree of polymerization \cite{BrownSzamel98,muller1996,Hur}
up to $N\lesssim10^{3}$. We briefly note that dropping the topological
excluded volume term in equation (\ref{eq:FET-1-1}) leads to $R\sim N^{1/4}$,
a result that formally corresponds to ideal hyperbranched polymers
and was found previously for non-concatenated rings in an array of
obstacles \cite{Obukhov1994}. 

Grosberg et al. \cite{Grosberg_Feigel_Rabin} and Sakaue \cite{Sakaue2011,Sakaue2012}discussed,
that the unknotting constraint leads to a free energy contribution
that turns into $\approx N^{3}/R^{6}$ for a large compression of
rings. This term results from the then dominating three body interactions
among the monomers of the ring and can be pictured as a simultaneous
compression in length and width of a the $R$-tube \cite{Grosberg_Feigel_Rabin}
that confines the ring. Such large compression is obtained above a
critical $N_{C}$ and adapting the third term of the free energy we
have 
\begin{equation}
\frac{F}{kT}\approx\frac{N}{R^{2}}+R^{2}+\frac{N^{3}}{R^{6}}.\label{eq:FET}
\end{equation}
Again, the entropy loss upon squeezing a ring appears to be of minor
importance for large $N$. The equilibrium extension of compressed
non-interpenetrating and unknotted rings with $N_{C}<N<N^{*}$ is
then given by 
\begin{equation}
R\sim N^{3/8}.\label{eq:sc1}
\end{equation}
A scaling of ring size with a power $<2/5$ for large $N>10^{3}$
is supported by computer simulation data as summarized, for instance,
in Fig. 1 of Ref \cite{Halverson}. This suggest that $N_{C}\approx10^{3}$
for the particular simulation models referenced in \cite{Halverson}.

To the best of our knowledge, there is no data available for $N\gg N^{*}$.
For such large molecular weights, the non-concatenation contribution
should be of the form as originally proposed by Cates and Deutsch
\cite{cates1986}, since then, all overlapping rings will be concatenated.
Balancing this expression with the dominating non-knotting contribution
for large $N$ yields $R\sim N^{4/9}.$ It will be interesting to
extend experimental or simulation data beyond $N^{*}$ in order to
test whether this proposed regime correctly describes the limiting
behaviour for $N\rightarrow\infty$. Furthermore we have to mention
that the available data suggests $N_{C}<N^{*}$ due to the observed
change of the scaling behaviour of rings in melts. However, $N_{C}<N^{*}$
is not necessarily a universal result, since $N^{*}$ depends on overlap
and the probability of trapping another ring, while $N_{C}$ is simply
defined by the onset of significant compression of the rings. Finally,
we would like to stress that the above Flory estimates have to be
taken with care and that there is no guarantee that the above exponents
can be considered as exact. But the general trend of whether compression
becomes weaker or stronger in the particular regimes might be correctly
reproduced.

Recently, Sakaue \cite{Sakaue2011,Sakaue2012} proposed an alternative
model for the size of rings in melt. The non-concatenation condition
was introduced similar to the original CD model, which is only correct
for sufficiently large $N>N^{*}$ according to our simulation data.
In contrast to the work of Sakaue, we find a well defined scaling
regime for ring size and not a gradual transition from a power 1/2
to 1/3. Furthermore, our approach does not require to postulate that
the topological length scale may be a function of the melt molecular
weight. Since both models lead to quite similar predictions for rings
size, it will be difficult to decide, which one migth be more appropriate
to describe conformations of rings in melt by discussing only the
size of rings in monodisperse melts. However, the differences between
both models should become most prominent in case of bimodal melts
of rings, which is, therefore, the subject of ongoing research.

\section{Summary}

In the present work, we compared the conformations of non-concatenated
entangled and interpenetrating disentangled melts of ring polymers.
It was found that the swelling of interpenetrating rings upon dilution
follows the same laws as for linear chains. Knotting and linking probabilities
of ring polymers in semi-dilute solution were analyzed using the HOMFLY
polynomial. We find an exponential decay of the non-knotting probability
of rings. The correlation length of the semi-dilute solution can be
used to superimpose knotting data at different concentrations. A power
law dependence $f_{n}\sim\phi R^{2}\sim\phi^{0.77}N$ for the average
number $f_{n}$ of linked rings per ring at concentrations larger
than the overlap volume fraction of rings $\phi^{*}$ is determined
from simulation data. The fraction of non-concatenated rings diplays
an exponential decay $P_{OO}\sim\exp(-f_{n})$, which indicates $f_{n}$
to provide the entropic effort for not forming concatenated conformations.
For very large overlap number $P$, it is expected that $f_{n}$ crosses
over to $f_{n}\sim P\sim\phi N^{1/2}$. The cross-over defines a lenght
scale $N^{*}$ which turns out to be very large, at least $N^{*}>1024$.
The exponential dependence of the non-linking probability as function
of $N$ in penetrating rings indicates a minimal degree of polymerization
$N_{OO}$, below which topology is unimportant. Furthermore, the available
simulation data \cite{Halverson} indicate a degree of polymerization
$N_{C}<N^{*}$ at which ring compression crosses over from weak at
$N<N_{C}$ to strong $N>N_{C}$.

Based upon these observations we discuss Flory-type free energy models
for the regimes separated by three cross-over degrees of polymerization,
$N_{OO}$, $N_{C}$, and $N^{*}$. For intermediate degrees of polymerization
$N_{OO}<N<N_{C}$ below significant compression of the rings, the
scaling of ring conformations might result from balancing non-concatenation
with the swelling term of ring topology. This leads to ring sizes
$R\sim N^{2/5}$. For larger $N>N_{C}$ but still below the transition
to the overlap dominated regime at $N^{*}$, significant compression
of the rings can occur, where non-concatenation is now balanced by
a free energy contribution upon strong compression $\approx N^{3}/R^{6}$,
which implies $R\sim N^{3/8}$. This slight decrease of the exponent
from $\text{\ensuremath{\approx}}2/5$, towards a smaller value close
to $1/3$ agrees with previous simulation results for larger rings
\cite{Halverson}. Finally, for extremely long rings with $N>N^{*}$,
a further change in the scaling behavior is expected. Here, all overlapping
pairs of rings are concatenated and the scaling of $f_{n}\sim R^{2}$
turns over into the scaling of the Flory number, $f_{n}\sim R^{3}/N$,
which may lead to ring size $R\sim N^{4/9}$ which is, however, rather
speculative in view of lacking simulation data.

Even though a large part of the discussion is supported by the data
presented in our work the major challenge is to fully explore the
limiting regime of very large $N>N^{*}$ and to determine $N^{*}$.
Furthermore, bidisperse blends of rings of different weight can serve
as critical test to clarify differences between the model presented
here and previous models \cite{cates1986,Sakaue2011,Sakaue2012} in
literature.

\section{Acknowledgement}

ML would like to thank the DFG under grant LA2375/2-1 and the ZIH
Dresden for a generous grant of computing time under the project BiBPoDiA.

\section*{\newpage{}}
\section*{Table of Contents Graphic}
\emph{Concatenation of rings}

Michael Lang

Jakob Fischer

Jens-Uwe Sommer

\begin{figure}[htbp]
\includegraphics[angle=0,width=\columnwidth]{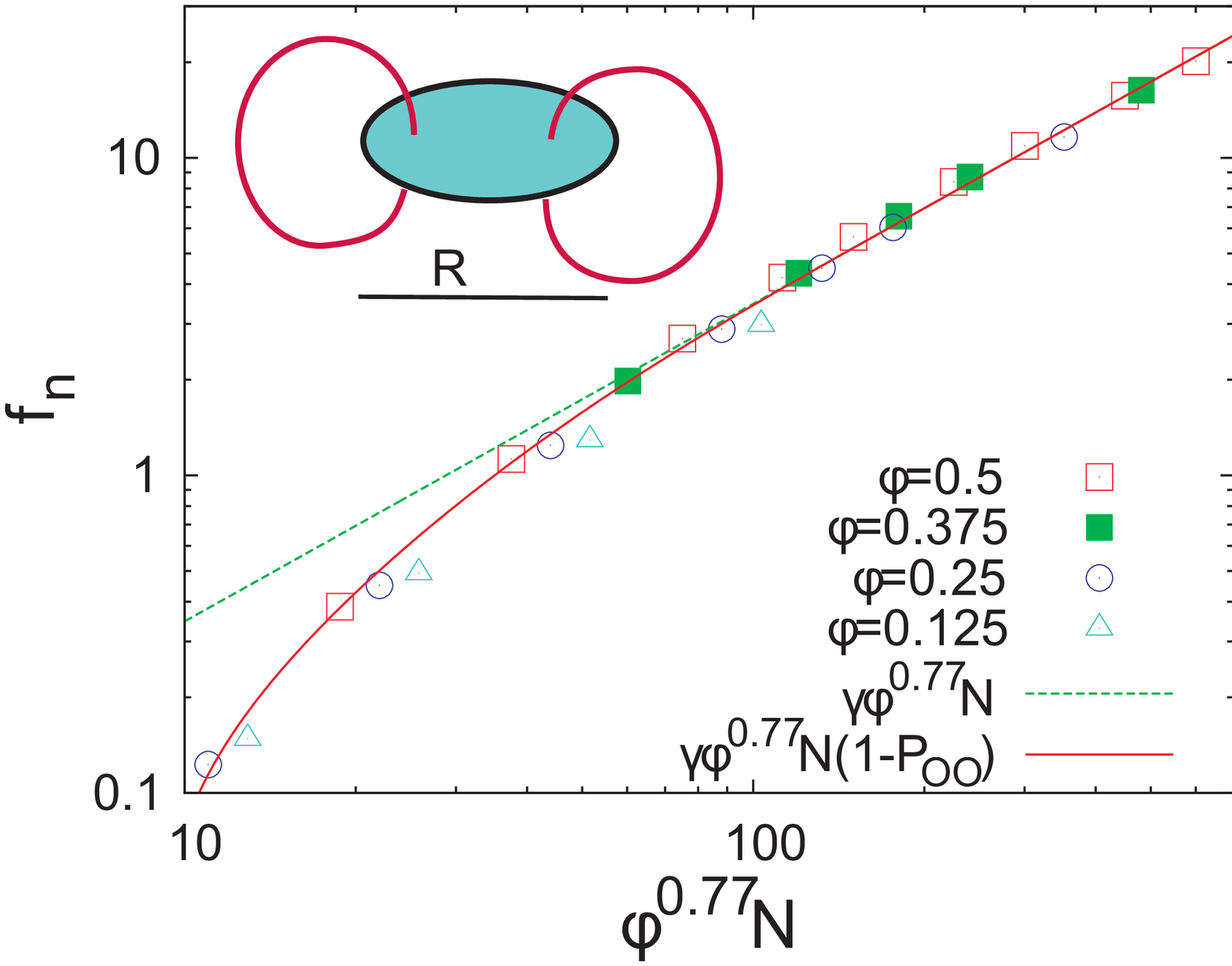}
\end{figure}
\end{document}